\documentclass[aps,pre,preprint]{revtex4-1}

\usepackage{amsbsy,amssymb,amsmath,bm}
\usepackage{graphicx,subfigure}
\usepackage{natbib}
\usepackage{color} 

\renewcommand{\Re}{\mathrm{Re}}
\newcommand{\Rm}{\mathrm{Rm}}
\newcommand{\VA}{V_{\textsc{a}}}

\begin{document}

\title{Numerical simulations of the Princeton magneto-rotational instability experiment with conducting axial boundaries}
\author{Xing Wei$^1$, Hantao Ji$^{1,2}$, Jeremy Goodman$^1$, Fatima Ebrahimi$^2$, Erik Gilson$^2$, Frank Jenko$^3$, Karl Lackner$^4$}
\affiliation{$^1$Princeton University Observatory \\ $^2$Princeton Plasma Physics Laboratory \\ $^3$UCLA Physics and Astronomy \\ $^4$Max Planck Institute for Plasma Physics}
\date{}

\begin{abstract}
We investigate numerically the Princeton magneto-rotational instability (MRI) experiment and the effect of conducting axial boundaries or endcaps. MRI is identified and found to reach a much higher saturation than for insulating endcaps. This is probably due to stronger driving of the base flow by the magnetically rather than viscously coupled boundaries. Although the computations are necessarily limited to lower Reynolds numbers ($\Re$) than their experimental counterparts, it appears that the saturation level becomes independent of $\Re$ when $\Re$ is sufficiently large, whereas it has been found previously to decrease roughly as $\Re^{-1/4}$ with insulating endcaps. The much higher saturation levels will allow for the first positive detection of MRI beyond its theoretical and numerical predictions.
\end{abstract}

\maketitle

\section{Introduction}

Magneto-rotational instability (MRI) was discovered by \citet{velikhov}, more systematically studied by \citet{chandrasekhar}, and applied to accretion disks by \citet{balbus}. Since the specific angular momentum ($\Omega r^2$) of such disks increases outward, they are hydrodynamically stable against Rayleigh's axisymmetric centrifugal instability. Numerical \cite{Hawley+etal1999,Lesur+Longaretti2005} and experimental \cite{ji,Edlund+Ji2014} evidence indicates that generic Keplerian and quasi-Keplerian flows are completely stable against purely hydrodynamic modes. It is therefore believed that the turbulence and angular-momentum transport in accretion disks is driven mainly by MRI \citep{balbus-review}. The Princeton MRI experiment has been designed to demonstrate the instability in a Taylor-Couette flow \cite{kageyama,nornberg,roach}. Relevant linear-stability analyses have been carried out at various levels of geometric fidelity \citep{goodman1,goodman2,Ruediger+etal2003}. Nonlinear calculations have been performed in axisymmetry with periodic boundary conditions \citep{liu} or insulating boundary conditions \citep{liu08} and in three dimensions with pseudo-vacuum boundary conditions \citep{gissinger}. The latter found that the saturation level of MRI decreases roughly as $\Re^{-1/4}$ with increasing Reynolds number ($\Re$) at fixed magnetic Reynolds number ($\Rm$). This can be explained heuristically by balancing viscous interaction at the boundaries against magnetic stresses in the bulk fluid. It was then considered that conducting endcaps may improve MRI. A linear boundary-layer calculation in the spirit of \citep{Gilman+Benton1968} but for a conducting boundary (see also \cite{siran,Szklarski+Ruediger2007}) gives the ratio of magnetic to viscous boundary-layer stresses as
\begin{equation}\label{eq:ratio}
\frac{\Gamma_{\rm mag}}{\Gamma_{\rm visc}}=\frac{\sqrt{2}\Lambda}{\left(1+\Lambda^2\right)^{1/4}}\left(\frac{\Omega\delta^2}{\nu}\right)^{1/2}.
\end{equation}
Here $\Lambda=\sigma_fB_0^2/\rho\Omega$ is an Elsasser number based on the fluid conductivity ($\sigma_f$) and density ($\rho$), the imposed field normal to the boundary ($B_0$), and the angular velocity $\Omega$, which is assumed to be slightly different for the boundary than for the fluid at large distances from it. The conductivity ($\sigma_s$) and geometrical thickness ($d_s$) of the boundary enter this expression via an effective electrical thickness $\delta=\sigma_sd_s/\sigma_f$. In the experimental parameter regime, the ratio~(\ref{eq:ratio}) is $\sim 10^3$, i.e. the magnetic coupling is 1000 times stronger than viscous coupling.

In this work, we use numerical calculations to test the effect of conducting endcaps on the MRI saturation level and other aspects of the flow. The basic equations are discussed in \S2, results in \S3, and experimental implications in \S4.

\section{Equations}

We solve the dimensionless Navier-Stokes and induction equations
\begin{equation}
\frac{\partial\bm u}{\partial t}+\bm u\cdot\bm\nabla\bm u=-\bm\nabla p+\frac{1}{\Re}\nabla^2\bm u+(\bm\nabla\times\bm B)\times\bm B\,,
\end{equation}
\begin{equation}
\frac{\partial\bm B}{\partial t}=\bm\nabla\times(\bm u\times\bm B)+\frac{1}{\sigma\Rm}\nabla^2\bm B\,,
\end{equation}
using cylindrical coordinates $(r,\theta,z)$ in a domain modeled after the Princeton MRI experimental setup, where the inner and outer cylinders have radii $r_1= 7$~cm and $r_2=$~21 cm, the height $h=$~28 cm, and the thickness of the newly installed copper endcaps $d_s=$~2 cm \citep{roach}. The magnetic induction equation is solved for both fluid and solid endcaps. A quasi-spherical vacuum region in which $\bm B = \bm\nabla\Psi$ surrounds the walls and endcaps, and on the outer boundary of this region we set the magnetic potential $\Psi=B_0z$. The radius of spherical vacuum is 10 times of the height of the cylindrical setup.

The computational units of length, time, magnetic field, and conductivity are $r_1$, $\Omega_1^{-1}$, $r_1\Omega_1\sqrt{\rho\mu_0}$, and $\sigma_f$ of the working liquid metal GaInSn, respectively. Here $\Omega_1$ is the angular velocity of the inner cylinder and $\rho$ is the density of the fluid. Note that all permeabilities have the vacuum value, $\mu_0$. The three dimensionless parameters governing this MHD system are the Reynolds number $\Re\equiv\Omega_1r_1^2/\nu$, the magnetic Reynolds number $\Rm\equiv\Omega_1r_1^2\sigma_f\mu_0$, and the Lehnert number $B_0\equiv \VA/\Omega_1r_1=\tilde B_0/\sqrt{\rho\mu_0}\Omega_1r_1$, where $\tilde B_0$ is the imposed field in dimensional units. Other dimensionless measures of the field strength can be expressed in terms of these: the Lundquist number $\mathrm{Lu}\equiv\VA r_1\mu_0\sigma_f= B_0\Rm$, the Elsasser number $\Lambda\equiv \VA^2/\eta\Omega=B_0^2\Rm$, and the Hartmann number $\mathrm{Ha}\equiv r_1\VA \sqrt{\mu_0\sigma_f/\nu}= B_0\sqrt{\Re\Rm}$. The experimental design limit for $\Omega_1$ is 4000 rpm, and so $\Re\sim O(10^7)$, $\Rm\sim O(10)$ and $B_0\sim O(0.1)$. The dimensionless thickness of the endcaps is $d_s=0.3$ (approximately), and their electrical conductivity $\sigma_s=19$.

In the regime where MRI occurs but centrifugal instability does not, it is required that $\Omega_1>\Omega_2$ but $r_1^2\Omega_1<r_2^2\Omega_2$: the ``quasi-Keplerian'' regime.  Inserting $r_2=3r_1$ we are led to the condition for MRI, $1/9<\Omega_2/\Omega_1<1$. In the calculations we take $\Omega_2/\Omega_1=0.1325$. To suppress Ekman (or Ekman-Hartmann) circulation driven by the boundary layers at the endcaps, both in these calculations and in the actual experiment, the endcaps are divided into two rings. No-slip conditions are applied at the boundaries, with angular velocities
\begin{align}\label{eq:bc}
\Omega_1&=1 & r&=1,\, -2\le z\le 2 \mbox{: inner cylinder};\nonumber\\
\Omega_3&=0.55 & 1&< r < 2,\, z=-2,2 \mbox{: inner ring};\nonumber\\
\Omega_2&=0.1325 & r&>2 \mbox{: outer ring ($z=0,h$) and cylinder ($r_2=3$)}.
\end{align}
The initial fluid velocity is piece-wise uniform rotation matched to the rings. The initial magnetic field is the imposed uniform vertical field $B_0$. Following \cite{gissinger}, departures from this initial field configuration are quantified by the volume-averaged radial field (``$B_r$ signal''):
\begin{equation}\label{eq:br}
\sqrt{\frac{1}{V}\int_V\left(\frac{B_r}{B_0}\right)^2dV}.
\end{equation}

The numerical calculations are carried out with the Spectral Finite Element Maxwell Navier-Stokes solver (SFEMaNS) \cite{guermond}.  A Fourier spectral method is used in azimuth ($\theta$), and finite elements in the meridional plane. In the experimentally accessible regime, MRI is expected to be axisymmetric, and so the calculations presented here are axisymmetric, although some non-axisymmetric calculations were made to test for shear layer instabilities \cite{wei,hollerbach,gissinger}.  Up to $18,000$ triangular finite elements were used in the meridional plane.

\section{Results}

The endcaps drive secondary circulation, so that MRI must be detected as a modification or bifurcation of the circulation rather than a linear instability \citep{gissinger}.  Calculations
at the experimental $\Re\sim 10^7$ would be prohibitive, so we begin by seeking MRI at $\Re=1000$ and later study trends up to $\Re=32,000$.

\begin{figure}
\centering
\subfigure[]{\includegraphics[scale=0.4]{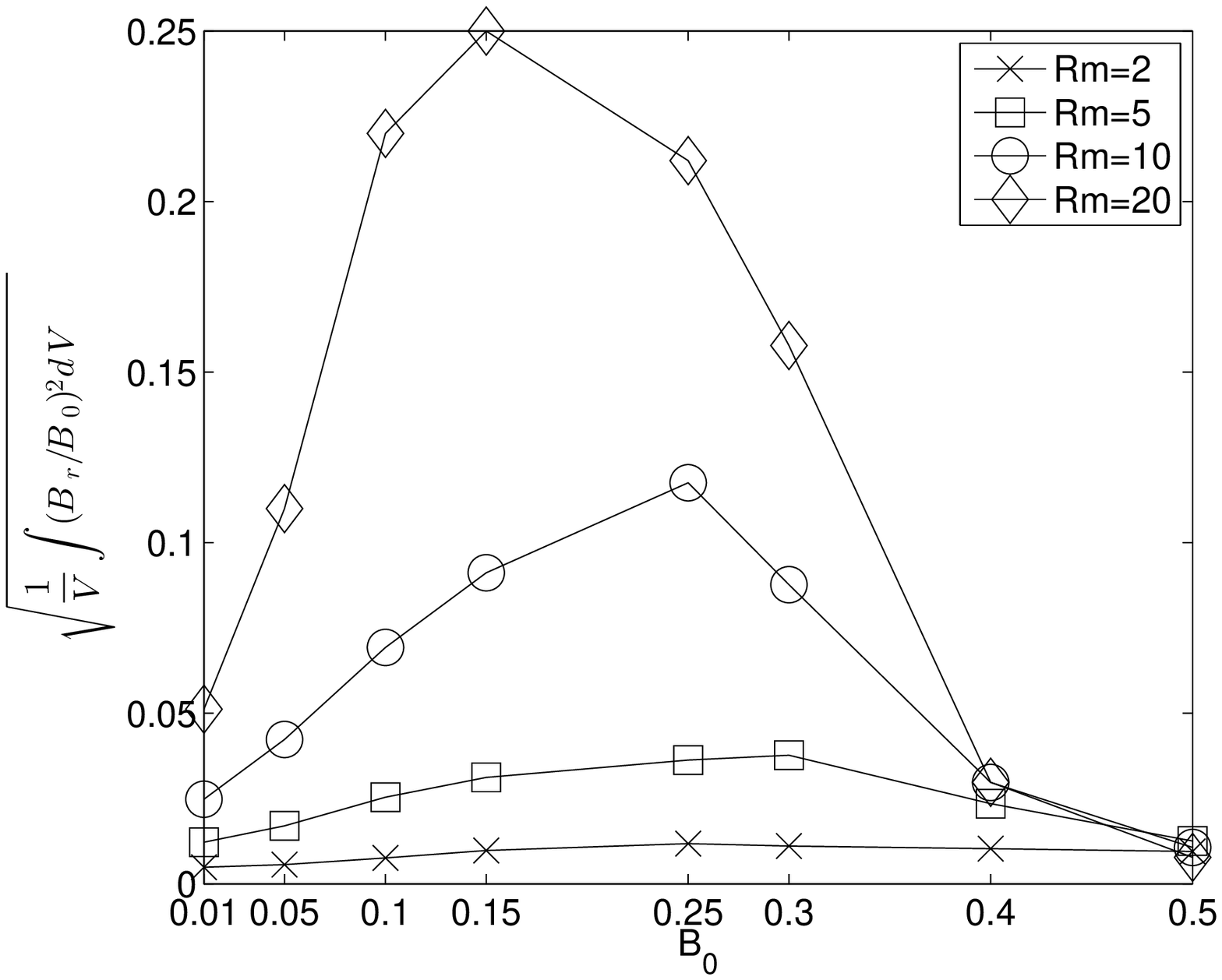}\label{fig1a}}
\subfigure[]{\includegraphics[scale=0.38]{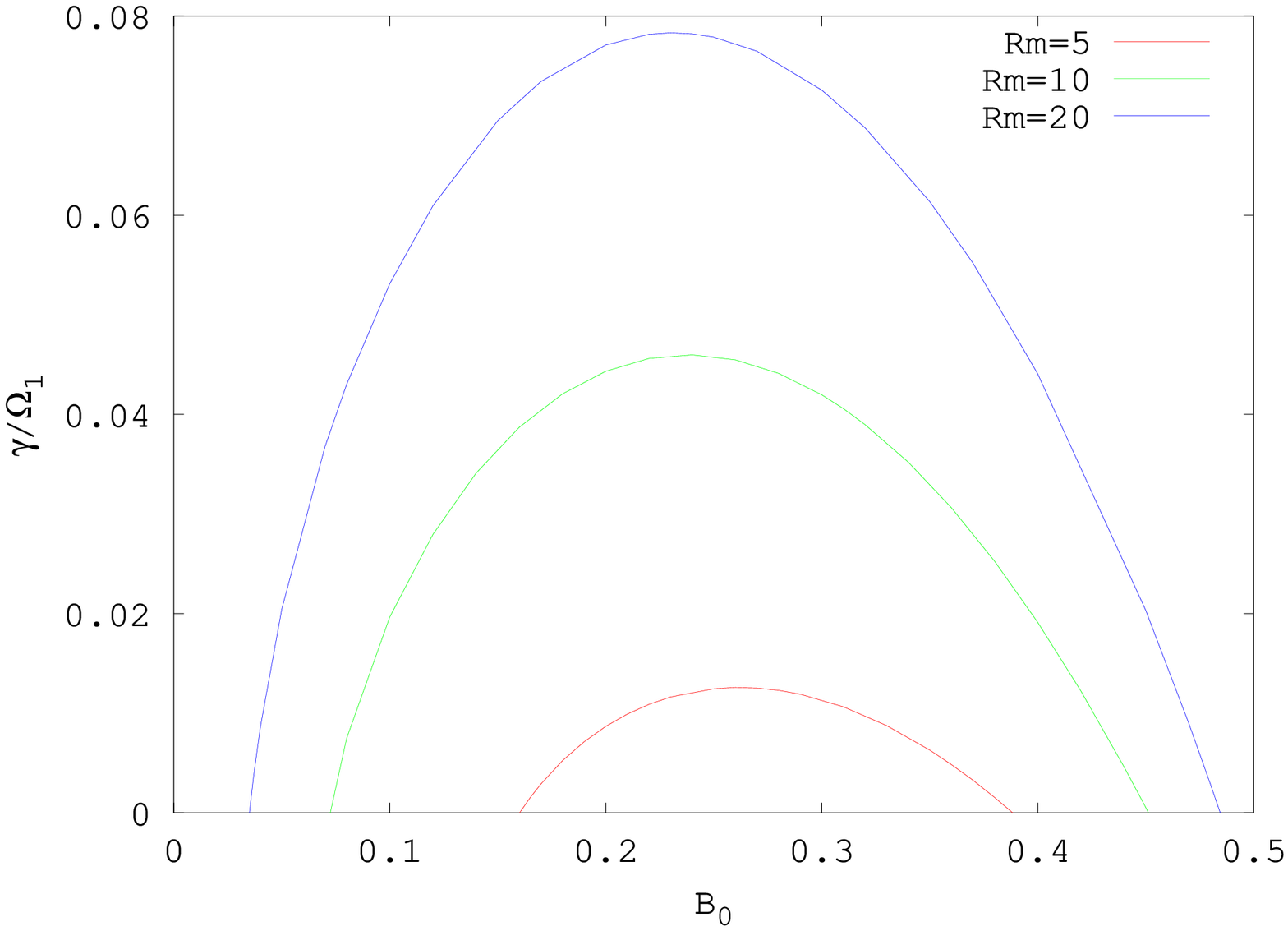}\label{fig1b}}
\subfigure[]{\includegraphics[scale=0.4]{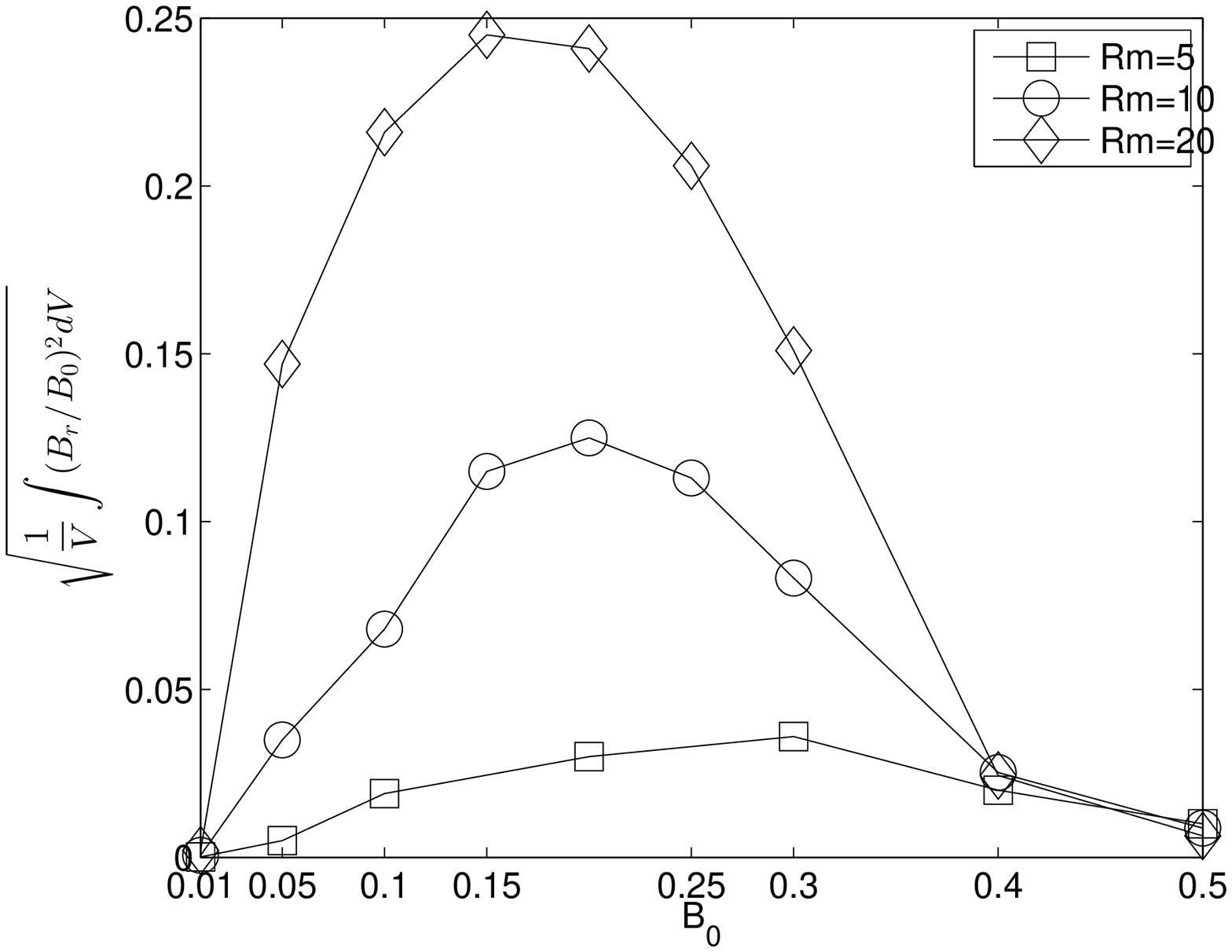}\label{fig1c}}
\subfigure[]{\includegraphics[scale=0.4]{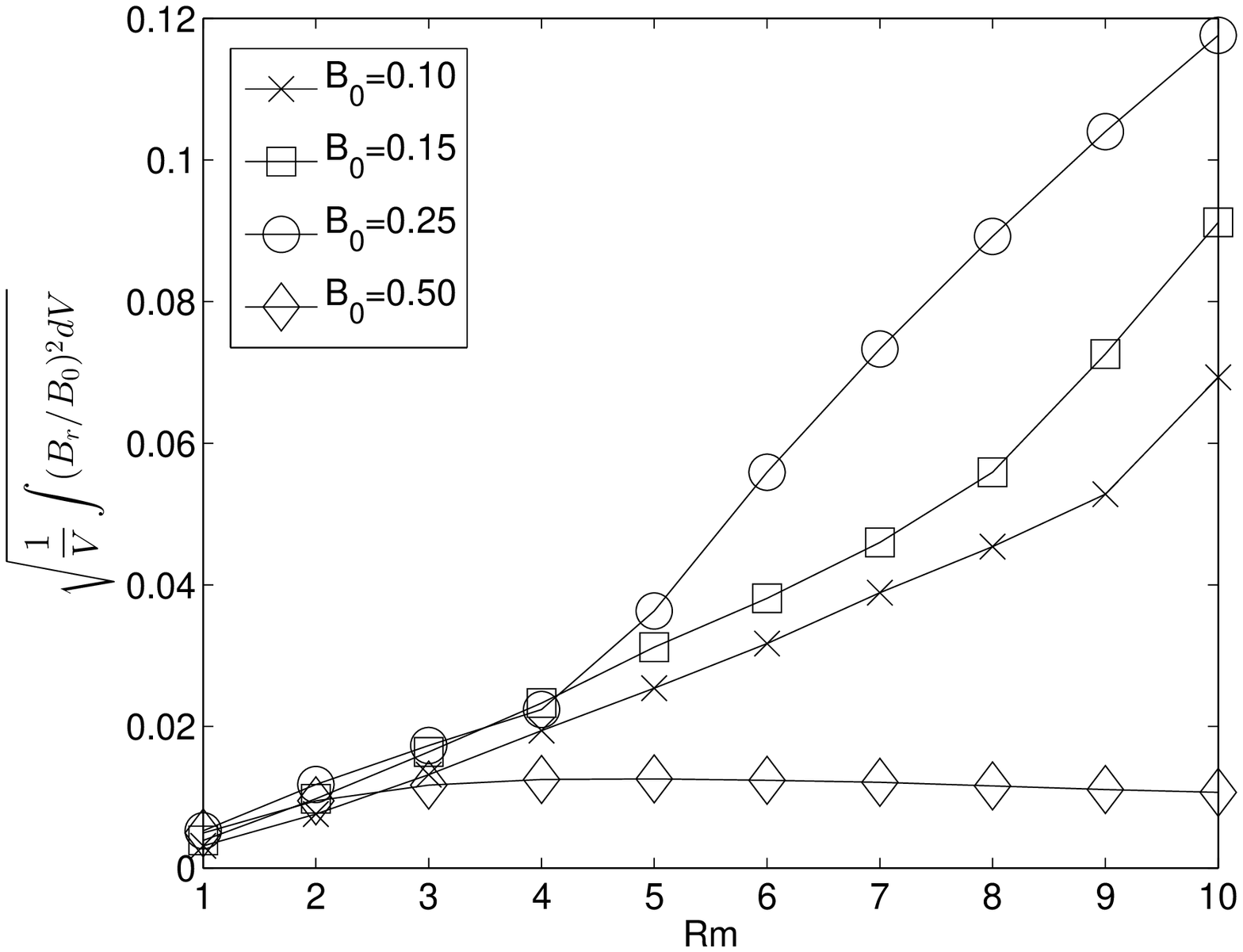}\label{fig1d}}
\caption{The $B_r$ signal at $\Re=1000$. (a) The $B_r$ signal versus $B_0$ for different $\Rm$,
   with conducting endcaps rotating as in eq.~(\ref{eq:bc}). (b)
  Linear growth rate of MRI versus $B_0$ for periodic vertical boundary conditions with period $2h$.
  (c) Like (a) but for insulating endcaps rotating with the ideal Taylor-Couette profile. (d)
  The $B_r$ signal versus $\Rm$ for different $B_0$, with endcaps as in (a).}\label{fig1}
\end{figure}

\begin{figure}
\centering
\subfigure[]{\includegraphics[scale=0.35]{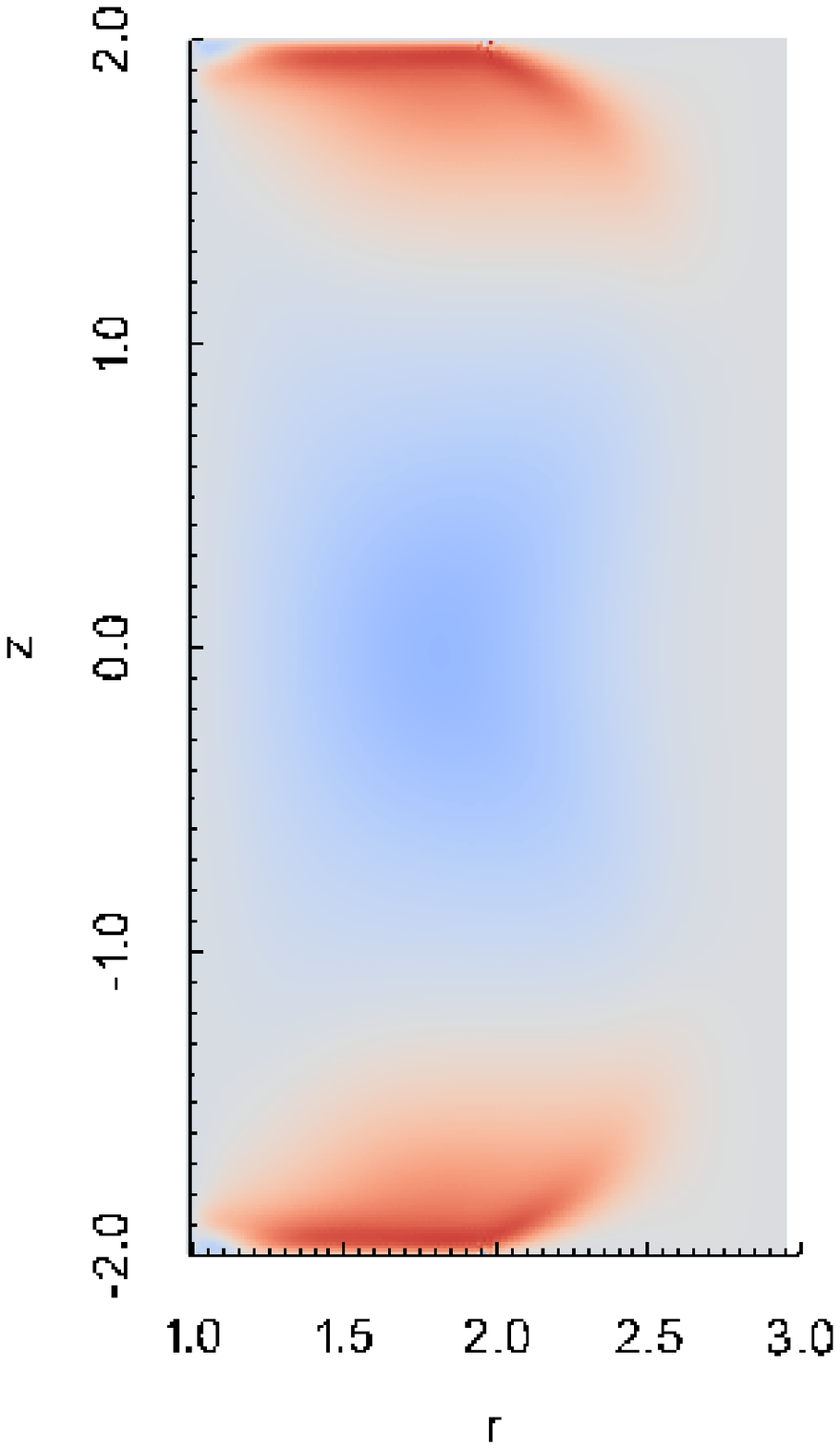}\label{fig2a}}
\subfigure[]{\includegraphics[scale=0.35]{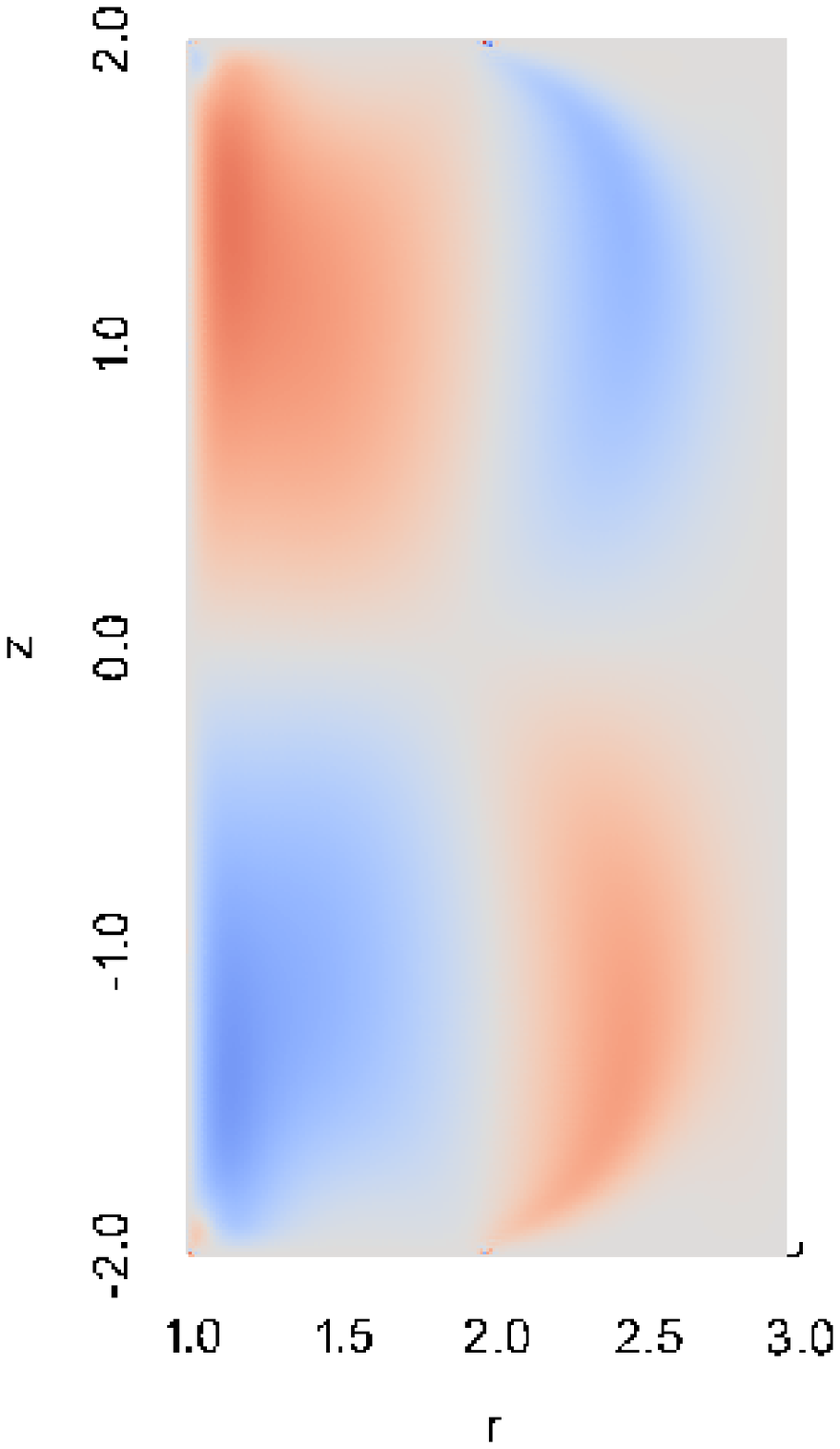}\label{fig2b}}
\subfigure[]{\includegraphics[scale=0.4]{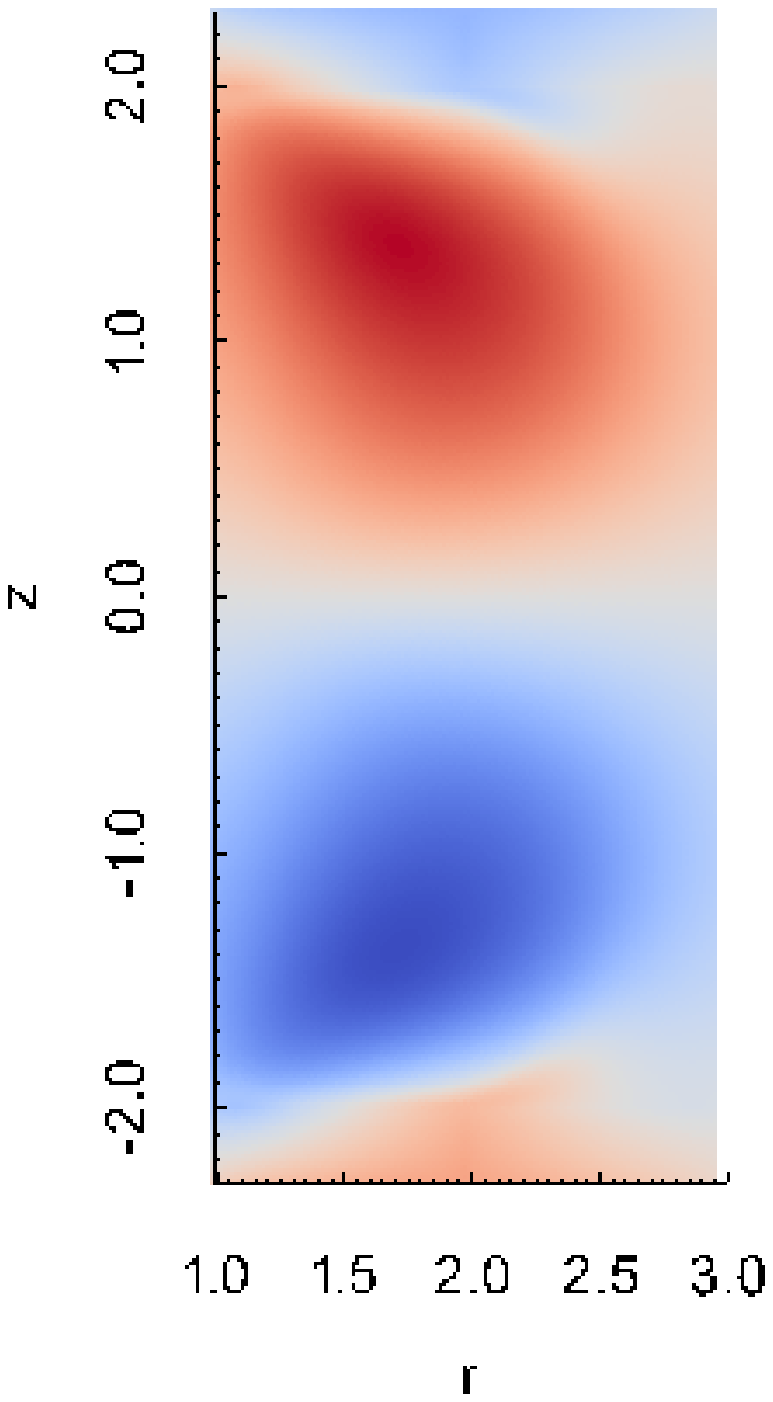}\label{fig2c}}
\subfigure[]{\includegraphics[scale=0.4]{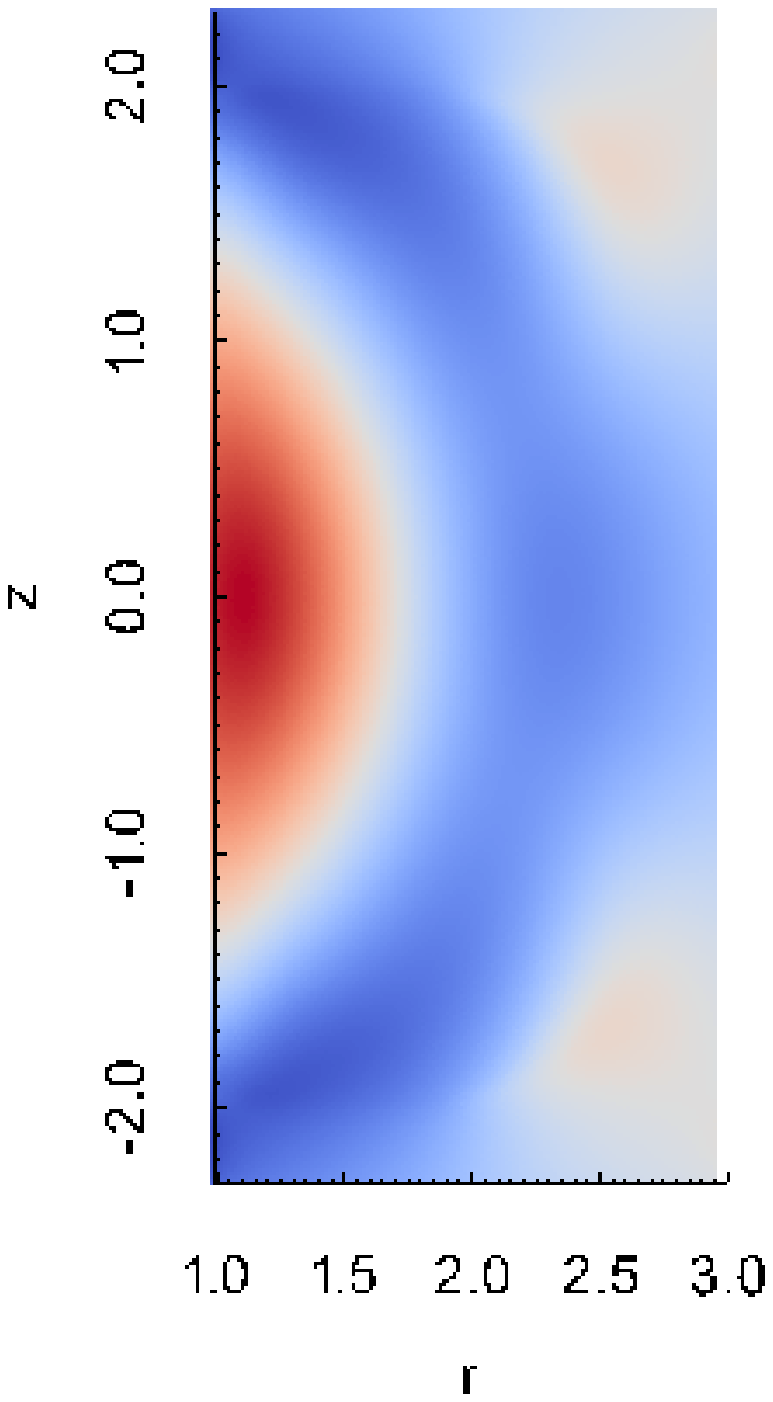}\label{fig2d}}
\caption{Meridional distributions.
(a) The radial velocity $u_r$. (b) The axial velocity $u_z$. (c) The radial field $B_r$. (d) The axial field $B_z$. Conducting endcaps at $\Re=1000$, $\Rm=20$ and $B_0=0.15$.}\label{fig2}
\end{figure}

Figure \ref{fig1a} shows the $B_r$ signal versus $B_0$ for different $\Rm$. The $B_r$ signal reaches
its maximum at the intermediate $B_0$ but is weak at both low and high $B_0$. This result is
consistent with the fact that MRI needs magnetic field but will be suppressed by a strong field
\cite{chandrasekhar}, and it is also consistent with the onset of MRI predicted by the local
analysis \cite{goodman1}, the global analysis \cite{goodman2} and the numerical calculation with
pseudo-vacuum boundary condition \cite{gissinger}. Figure \ref{fig1b} shows the linear growth rate
versus $B_0$ predicted by the methods of \cite{goodman2} for the same $\Re$ in vertically periodic
cylinders with vertical wavelength $2h$, which approximates the magnetic geometry at saturation
(Fig~\ref{fig2}). Evidently, the $B_r$ signal at saturation and the expected MRI linear growth rate
have similar dependence on the field strength. 

To suppress meridional circulation and isolate MRI signatures, we have also performed
  simulations in which the rotation of the insulating endcaps follows the ideal Taylor-Couette flow profile,
  $\Omega(r)=a+b/r^2$.  Although not feasible experimentally, such differentially rotating endcaps
  would permit a basic state of purely azimuthal motion following the ideal profile at all heights, and
  deviations could be interpreted as evidence for MRI (or perhaps other
  instabilities) rather than Ekman circulation.  Figure \ref{fig1c} shows the $B_r$ signal versus
  $B_0$ for these simulations.
Note that the $B_r$ signal vanishes as $B_0\to0$, unlike Fig.~\ref{fig1a}, as one expects in the
absence of Ekman circulation.  Apart from this,
the general similarity of panels \ref{fig1a} and \ref{fig1c} suggests that the $B_r$ signal in
  both cases is dominated by MRI, or at least not by meridional circulation of the basic state.
We also tested conducting endcaps with the smoothly varying rotational profile of ideal Taylor-Couette flow, and the results are slightly different from the previous insulating endcaps (the difference arises from the numerical error of the code). This suggests that the higher saturation levels of MRI with conducting endcaps may be due to the stronger driving of the base flow by conducting endcaps.

Figure \ref{fig1d} shows the dependence of $B_r$ on $\Rm$ at several $B_0$. For the three weaker fields, the variation with $Rm$ is monotonic but changes slope at an $\Rm$ that itself decreases with $B_0$: $\Rm\approx 9$ for $B_0=0.10$, at $\Rm\approx 8$ for $B_0=0.15$ and at $\Rm\approx4$ for $B_0=0.25$.
It is known that $B_r$ can be induced by either the Ekman-Hartmann circulation or MRI, and moreover, in different parameter regimes the circulation-induced $B_r$ and the MRI-induced $B_r$ may depend differently on the dimensionless parameters. At least in uniform rotation, the thickness of the boundary layer and the mass flux through it decrease monotonically with increasing Elsasser number ($B_0^2\Rm$) at fixed $\Re$ \citep{Gilman+Benton1968,Szklarski+Ruediger2007}, and the $B_r$ signal behaves similarly. MRI, on the other hand, grows fastest at intermediate $B_0$, as Fig.~\ref{fig1b} illustrates. Therefore, we take the data in Figure~1 as evidence for the onset of MRI.

Figure \ref{fig2} shows the meridional distributions of $u_r$, $u_z$, $B_r$ and $B_z$ at $\Re=1000$,
$\Rm=20$ and $B_0=0.15$, which is well within the MRI regime as discussed above, i.e. $\Rm>\Rm_c=8$
for $B_0=0.15$ as shown by figure \ref{fig1d}. Figure \ref{fig2a} shows that the radial flow mainly
lies in the boundary layer. Figure \ref{fig2b} shows the pumping arising from the Ekman-Hartmann
boundary layer. The combination of figures \ref{fig2a} and \ref{fig2b} shows the pattern of
circulation, i.e. in two opposing cells, clockwise at $z>0$ and counterclockwise at $z<0$. Figure
\ref{fig2c} shows that $B_r$ varies almost monotonically with height and the strongest $B_r$ appears
at $z\approx\pm1.5$ away from the boundary layer where the strongest $u_r$ appears. The different
locations of the strongest $B_r$ and $u_r$ suggest that $B_r$ is mainly induced not by circulation
(i.e. interaction of $u_r$ and $B_0$) but by MRI. Figure \ref{fig2d} shows that the strongest $B_z$
appears at the mid plane $z=0$ where $u_z$ is almost zero. Again, this suggests that $B_z$ is mainly
induced not by circulation (i.e. interaction of $u_z$ and $B_r$) but by MRI. 

\begin{figure}
\centering
\subfigure[]{\includegraphics[scale=0.4]{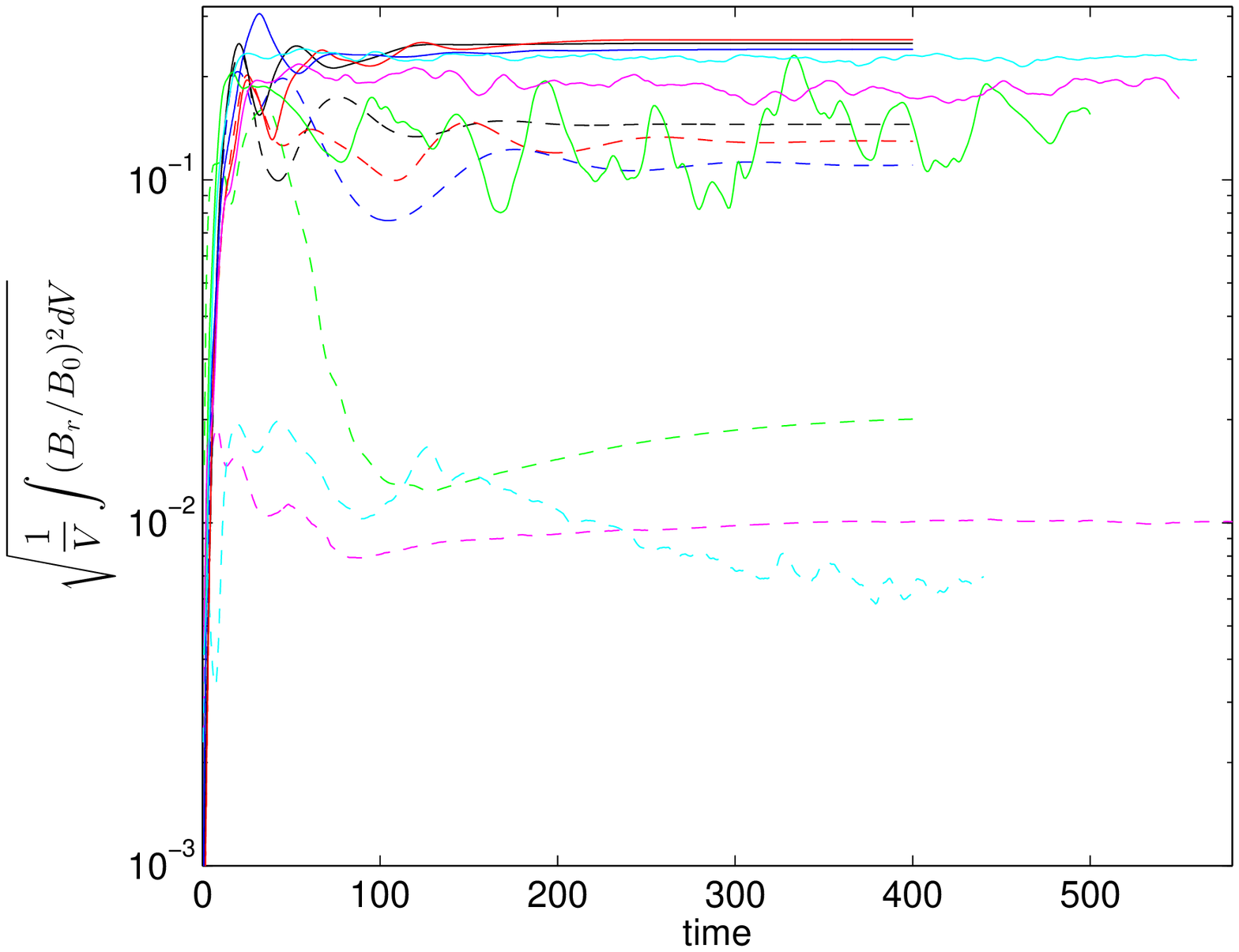}\label{fig3a}}
\subfigure[]{\includegraphics[scale=0.4]{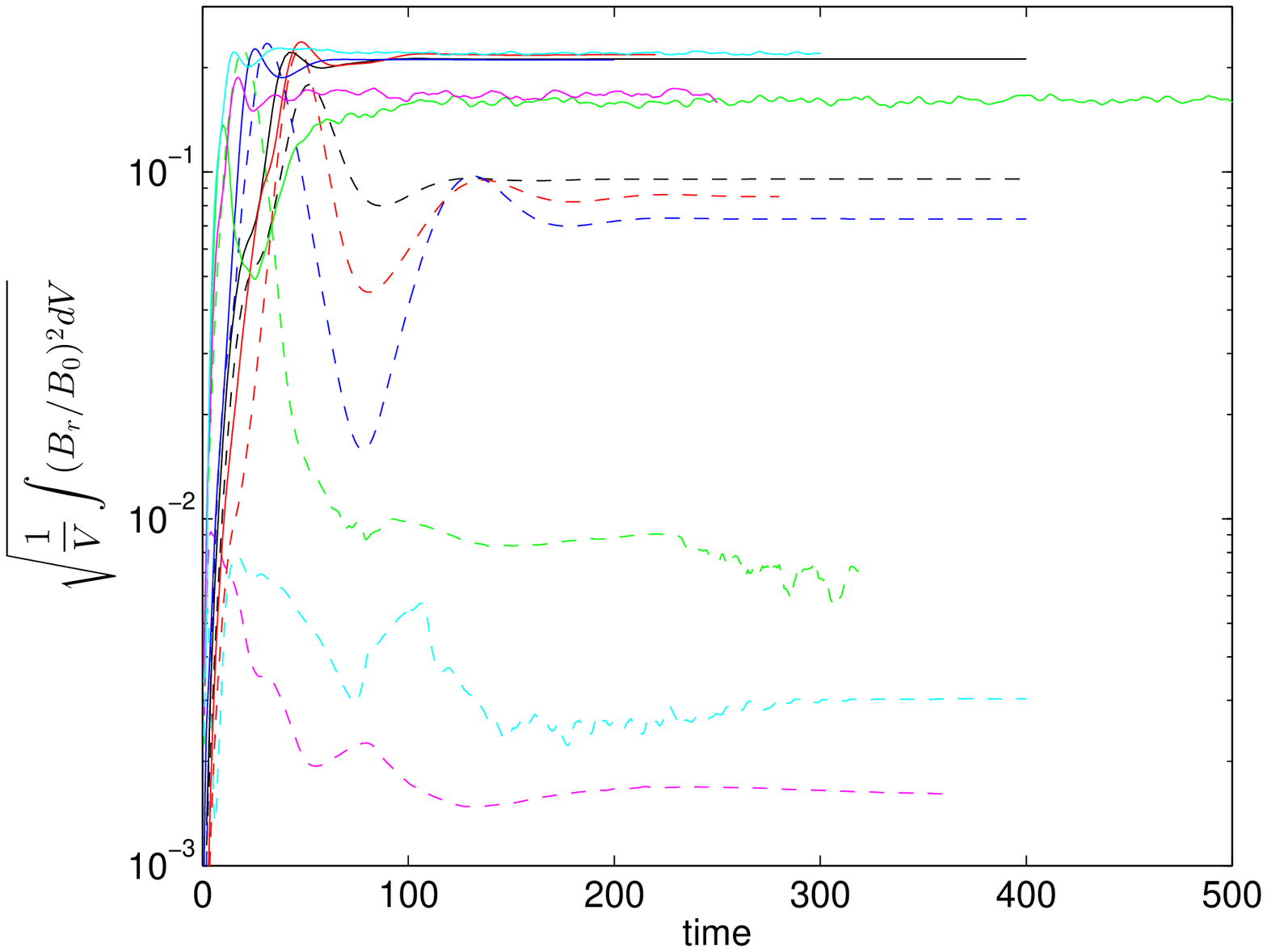}\label{fig3b}}
\caption{Conducting and insulating endcaps at $\Rm=20$. (a) The $B_r$ signal versus time for different $\Re$ at $B_0=0.15$. (b) The $B_r$ signal versus time for different $\Re$ at $B_0=0.25$. Solid lines denote conducting endcaps and dashed lines denote insulating endcaps. Black, red, blue, green, magenta and cyan colors correspond respectively to $\Re=1000$, $2000$, $4000$, $8000$, $16000$ and $32000$. Time unit is $\Omega_1^{-1}$.}\label{fig3}
\end{figure}

\begin{figure}
\centering
\includegraphics[scale=0.4]{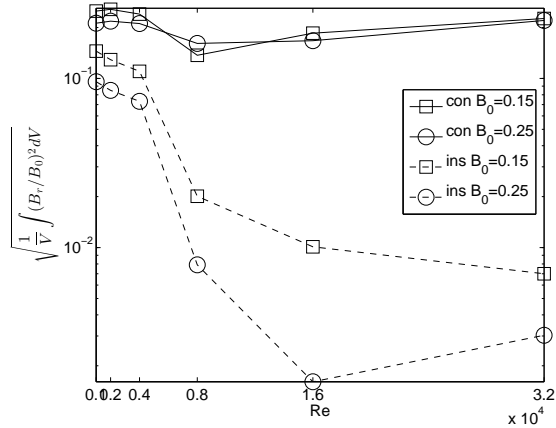}
\caption{The $B_r$ signal at $\Rm=20$ versus $\Re$ for different $B_0$ and magnetic boundary conditions, as in figure \ref{fig3}.}\label{fig4}
\end{figure}

We have also made calculations at higher $\Re$, in order to extrapolate toward the experimental regime. Figure \ref{fig3} shows the time evolution of the $B_r$ signal at $\Rm=20$ for different $\Re$ and magnetic boundary conditions. The left and right panels are for respectively $B_0=0.15$ and $0.25$, and both of them are in the MRI regime. With insulating endcaps, the $B_r$ signal becomes almost time-independent after the initial transient.
 With conducting endcaps, the $B_r$ signal fluctuates for $\Re\ge8000$. The variation of the time-averaged $B_r$ signal with $\Re$ is shown in Figure~\ref{fig4}. Clearly, the $B_r$ signal is higher with conducting than with insulating endcaps, and the contrast increases with increasing $\Re$ (Fig.~\ref{fig4}). At the highest $\Re=3.2\times10^4$, the $B_r$ signal with conducting endcaps is around $30$ ($70$) times that with insulating endcaps for $B_0=0.15$ ($0.25$). This is qualitatively consistent with eq.~\eqref{eq:ratio}. The $B_r$ signal scales differently with $\Re$ for the two boundary conditions. With insulating endcaps the $B_r$ signal decreases with increasing $\Re$, though more slowly above $\Re=8000$. With the conducting endcaps, the signal is approximately constant at large $\Re$, as might be expected if the flow is sustained mainly by magnetic rather than viscous coupling to the boundaries.

\section{Discussion}

In this work we carried out the preliminary numerical calculations for the Princeton MRI experiment to test the effect of conducting (copper) endcaps. Although the fluid Reynolds number of the numerical calculations is far away from the experimental regime, $\Re=O(10^7)$, these calculations have guided the redesign of the experiment. Firstly, MRI has been identified in a regime of intermediate $B_0$ and modest $\Rm$. According to these calculations, with conducting endcaps, MRI is very likely to be discovered at $\Rm\ge5$
 and $B_0\approx0.2$-$0.3$. 
In the experiment, the maximum $\Rm$ reaches 9 and $B_0$ can be 
selected within the range 0-0.3.
Secondly, conducting endcaps greatly increase the MRI signal, to a level $\sim 20\%$ at experimentally accessible $\Rm$ and $B_0$. This is sufficiently strong to be detected by the Hall probe newly installed on the experimental setup.

The discontinuous rotation profile (\ref{eq:bc}), if it extends from the endcaps into
  the fluid (forming a so-called Shercliff layer), might be expected to excite nonaxisymmetric instabilities of the
  Kelvin-Helmholtz type \cite{wei,roach}.  To test this, nonaxisymmetric simulations were performed
  with azimuthal wavenumbers up to $m=4$ and Reynolds numbers up to $Re=32,000$.  Negligible energy
  was found in the non-axisymmetric components ($m>0$).  This is in accord with the experimental
  results of \cite{roach}, who found that Shercliff-layer instabilities grow robustly in this
  apparatus (but before the endcaps were made conducting) only when the Elsasser number
  $\Lambda=\sigma \tilde B_0^2/\rho\Omega$ is $\ge 1$, whereas $\Lambda\lesssim1$ in
  the simulations of this paper.

\begin{figure}
\centering
\includegraphics[scale=0.5]{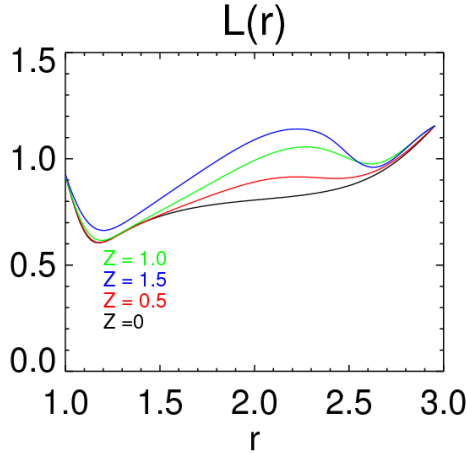}
\caption{The radial profile of specific angular momentum $L\equiv r^2\Omega$ at several heights, as
  marked.  
Other parameters ($\Re$, $\Rm$, $B_0$) as in Fig.~\ref{fig2}.}\label{fig5}
\end{figure}

Also possibly relevant is Rayleigh's centrifugal instability, which may arise where
  the specific angular momentum decreases outward, $\partial |r^2\Omega|/\partial r<0$.  We plot the
  radial angular-momentum profile in Figure~\ref{fig5} for the same parameters as in
  Figure~\ref{fig2}. Evidently, the angular momentum increases radially outward except near the
  inner cylinder, $1.0<r<1.2$, and at $2.3\lesssim r\lesssim 2.6$.  According to Figure \ref{fig2c}
  however, $B_r$ is largest at $1.5\lesssim r\lesssim 2.0$, where the flow is locally centrifugally
  stable ($d|L|/dr>0$).  In short, it seems unlikely that the $B_r$ signal is dominated by
  Shercliff-layer or Rayleigh instabilities.

To summarize, we have the following evidence for the presence of MRI in our simulations:
\begin{itemize}
\item The similarity of the normalized $B_r$ signal at nonlinear saturation to the linear MRI growth
  rate---in particular, the monotonic dependence on magnetic Reynolds number $Rm$ and non-monotonic dependence on 
the background field $B_0$ [Fig~\ref{fig1b}];
\item The similarity of the signal to that obtained in simulations where Ekman simulation is
  suppressed  [Fig~\ref{fig1c}];
\item A change in the slope of the dependence of the signal on $Rm$ at the predicted threshold of MRI  (Fig~\ref{fig1d});
\item Different spatial distributions of radial velocity $u_r$ (largest near boundaries) and
  radial field $B_r$ (largest in the bulk) [Fig.~\ref{fig2}];
\item Absence of non-axisymmetric modes;
\item Lack of spatial correlation of the signal with conditions favoring centrifugal instability.
\end{itemize}

\begin{figure}
\centering
\subfigure[]{\includegraphics[scale=0.4]{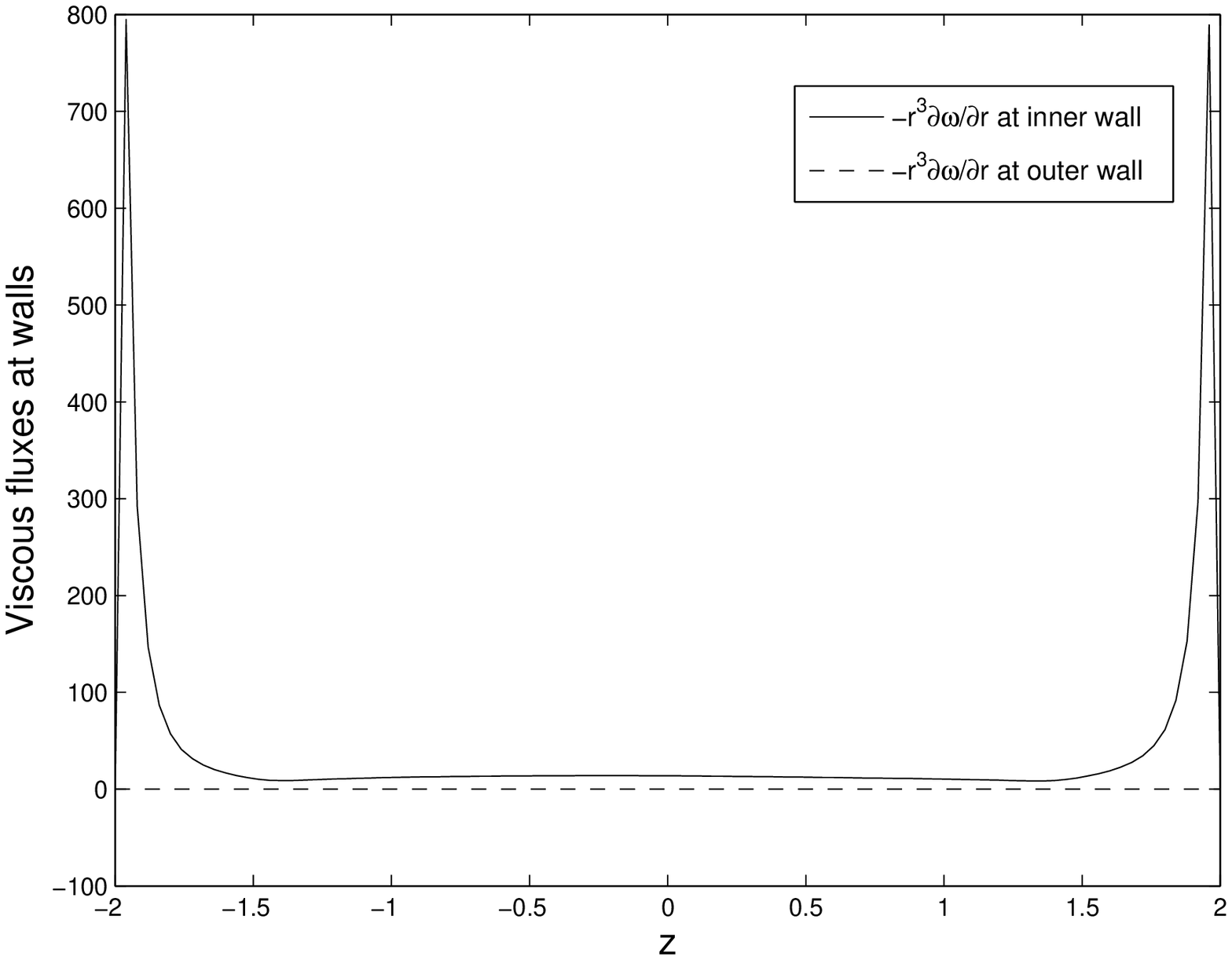}\label{fig6a}}
\subfigure[]{\includegraphics[scale=0.4]{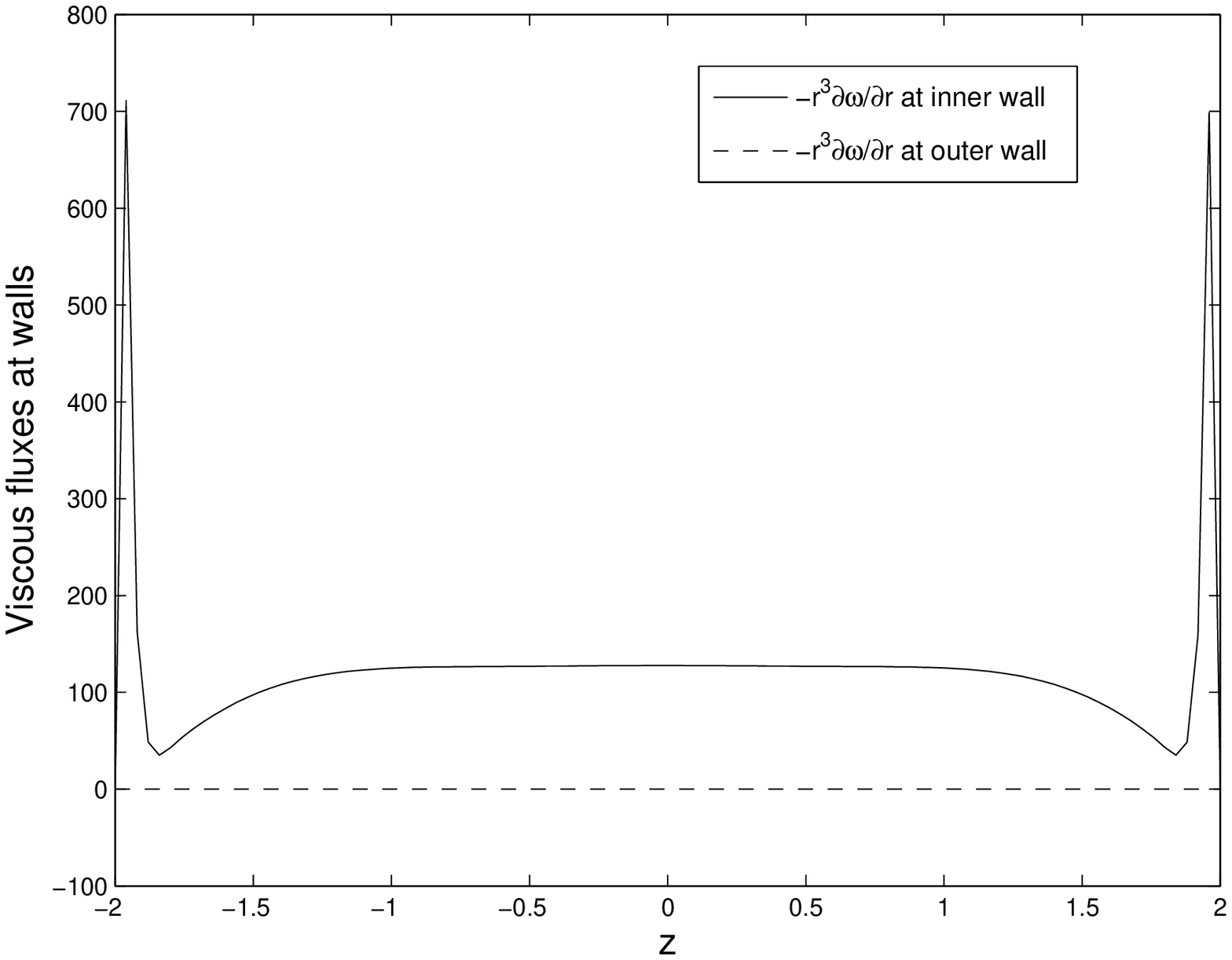}\label{fig6b}}
\subfigure[]{\includegraphics[scale=0.4]{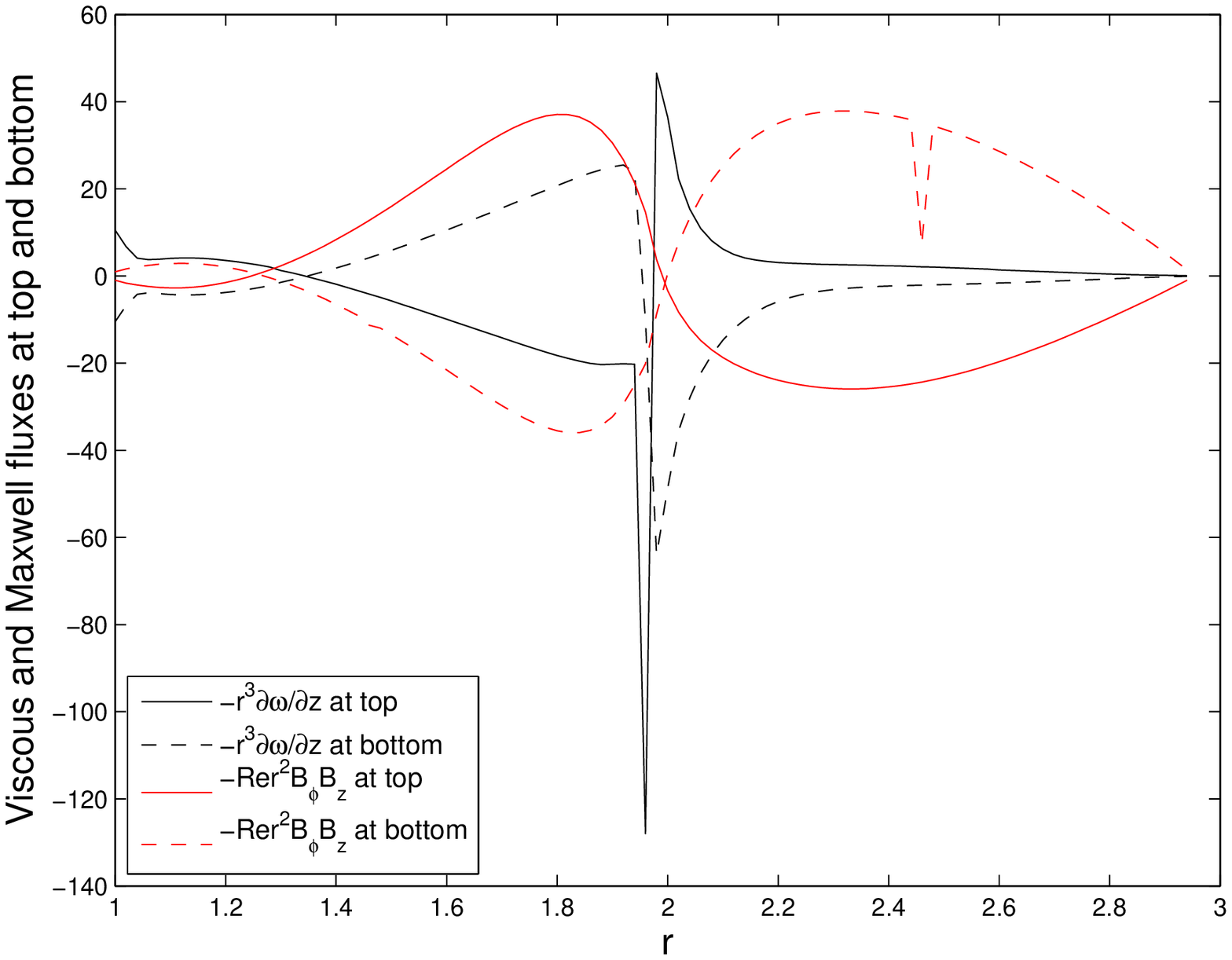}\label{fig6c}}
\subfigure[]{\includegraphics[scale=0.4]{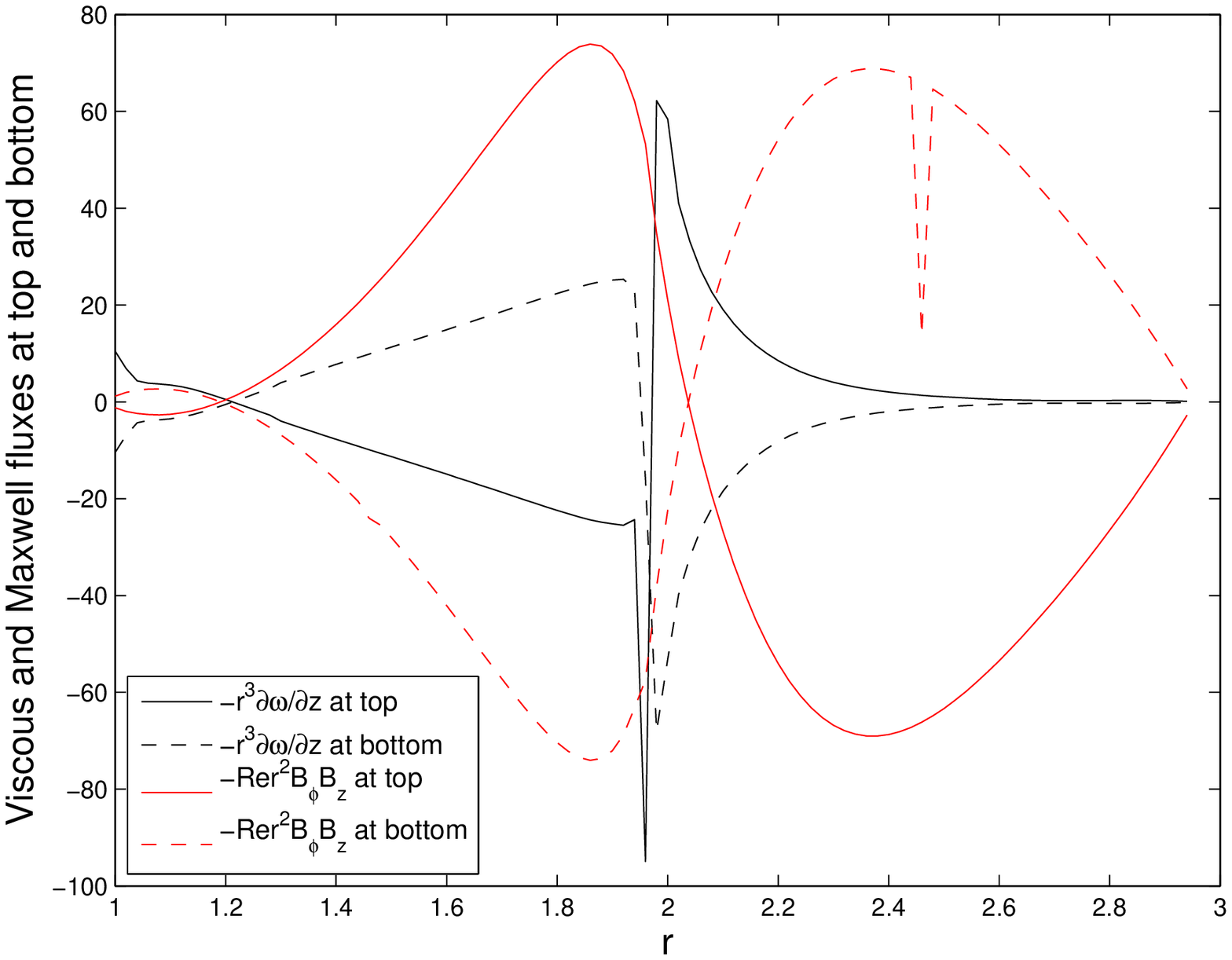}\label{fig6d}}
\caption{Viscous and Maxwell fluxes. (a) and (b): radial viscous fluxes $-r^3\partial\omega/\partial r$ at the inner and outer walls. (c) and (d): axial viscous fluxes $-r^3\partial\omega/\partial z$ and Maxwell fluxes $-\Re r^2 B_\phi B_z$ at the top and bottom. $Rm=10$ for (a) and (c) and $Rm=20$ for (b) and (d). $\Re=1000$ and $B_0=0.15$.}\label{fig6}
\end{figure}

To end this paper we briefly discuss the flux of axial angular momentum.
In the axisymmetric case, the radial and axial components of this flux are
\begin{align}\label{eq:flux}
F_r&=\rho r\left(u_ru_\phi-\frac{B_rB_\phi}{\rho\mu_0}-\nu r\frac{\partial\omega}{\partial r}\right), \nonumber\\
F_z&=\rho r\left(u_\phi u_z-\frac{B_\phi B_z}{\rho\mu_0}-\nu r\frac{\partial\omega}{\partial z}\right),
\end{align}
where $\omega=u_\phi/r$ is the angular velocity. In a steady state, or in the time average, $\bm\nabla\cdot\bm F=0$, and the flux integrated over the boundaries should vanish. Since the Reynolds stresses $u_ru_\phi$ and $u_\phi u_z$ vanish at the boundaries, we compare the viscous and Maxwell fluxes. We normalize the stresses with $\nu\Omega_1$ such that a pre-factor $\Re$ appears in the dimensionless expression of the Maxwell fluxes. We are concerned with the viscous and Maxwell fluxes across the boundaries, namely $2\pi rF_r$ versus $z$ at $r=r_1$ and $r_2$ and $2\pi rF_z$ versus $r$ at $z=\pm h/2$. Figure \ref{fig6} shows the radial fluxes at the inner and outer walls (Figs.~\ref{fig6a}, \ref{fig6b}), and the axial fluxes at the top and bottom endcaps (Figs.~\ref{fig6c}, \ref{fig6d}) at $Rm=10$ and $20$. For the radial fluxes, the Maxwell fluxes vanish because the walls are insulating and the viscous flux at the inner wall dominates over the one at the outer wall. At the endcaps, the magnetic flux is lower than the viscous at the low $Rm=10$ but exceeds the viscous at the large $Rm=20$. Figures \ref{fig6c} and \ref{fig6d} show that both the viscous and Maxwell fluxes change their signs between the inner and outer rings. The integral of the difference between the top and bottom endcaps is greater than that between the outer and inner walls, and thus the transport of angular momentum is substantially axial rather than purely radial. This is not what is usually envisaged for MRI transport in accretion disks. However, in the disks of protostars (and perhaps other disks, including those of quasars), much of the angular momentum may be removed via magnetic stresses that couple to an outflow (magneto-centrifugal wind), and in some parameter regimes this may be accompanied by MRI turbulence within the disk \cite{SKW2007b}.

\begin{acknowledgments}
Prof. Guermond provided us the SFEMaNS code, and Dr Cappanera and Prof. Nore helped use this code. One anonymous referee suggested to do calculations of ideal Taylor-Couette flow for a better evidence of MRI, and the other referee suggested to compare between insulating and conducting endcaps with ideal Taylor-Couette flow. This work was supported by the National Science Foundation's Center for Magnetic Self-Organization under grant PHY-0821899, the NSF grant AST-1312463, and the NASA grant NNH15AB25I. We thank the Max-Planck-Princeton Center for Plasma Physics (MPPC) for the collaboration of FJ and KL.
\end{acknowledgments}

\bibliographystyle{apsrev4-1}
\bibliography{paper}

\end{document}